\documentstyle[epsfig]{aipproc}

\newcommand{\bm}[1]{\mbox{\boldmath $#1$}}

\newcommand{\bkt}{\bm k_T^{}}
\newcommand{\bSt}{\bm S_T}
\newcommand{\ba}{\begin{eqnarray}}
\newcommand{\ea}{\end{eqnarray}}
\newcommand{\beq}{\begin{equation}}
\newcommand{\eeq}{\end{equation}}

\newcommand{\simorder}{\raisebox{-4pt}{$\, \stackrel{\textstyle >}{\sim} \,$}}

\begin{document}
\title{Transversely polarized $\Lambda$ production\footnote{Contribution to
the proceedings of the 7th Conference on the Intersections of Particle and 
Nuclear Physics (CIPANP 2000), Quebec City, Canada, May 22 - 28, 2000}}

\author{Dani\"el Boer\\[-5 mm]}
\address{RIKEN-BNL Research Center\\
Brookhaven National Laboratory, Upton, NY 11973, U.S.A.\\[-5 mm]}

\maketitle

\begin{abstract}
Transversely polarized $\Lambda$ production in hard 
scattering processes is discussed in terms of a 
leading twist T-odd fragmentation function which describes
the fragmentation of an unpolarized 
quark into a transversely polarized $\Lambda$.
We focus on the properties of this function and its relevance for the 
RHIC and HERMES experiments. 
\end{abstract}

\section*{Introduction}

Transverse polarization distribution and 
fragmentation functions parameterize transverse spin effects in hard
scattering processes. The question is which of these functions might 
be relevant for the description of the single transverse spin 
asymmetry in the process 
$p \, p \rightarrow \Lambda^{\uparrow} \, X$ \cite{Lambdadata}?
The fact that in this asymmetry the
transverse spin and the transverse momentum appear to be correlated --they 
are orthogonal to each other--, indicates that a so-called T-odd function is
required.

\section*{T-odd fragmentation functions}

\begin{figure}[htb] 
\centerline{\epsfig{file=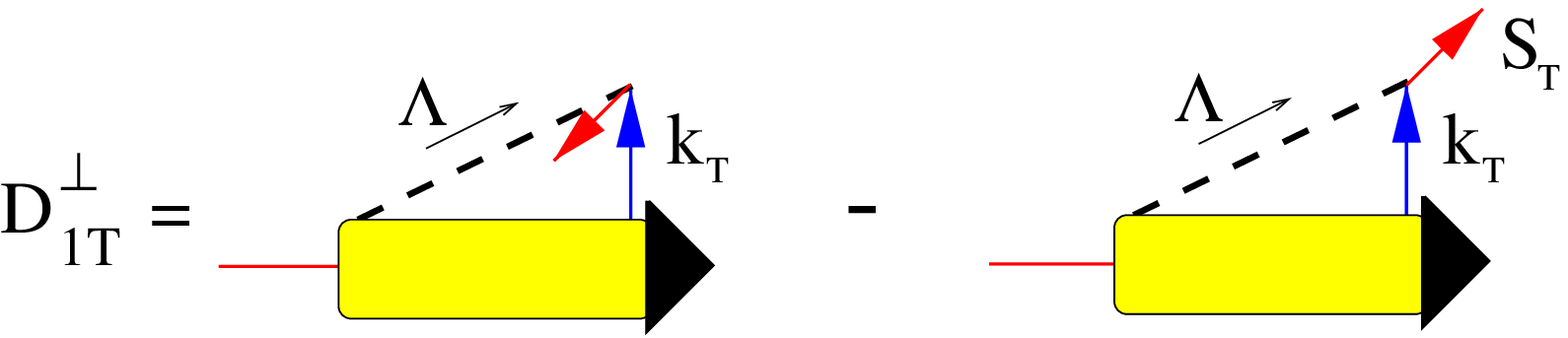,width=4in}}
\vspace{10pt}
\caption{The 
function $D_{1T}^\perp$ signals different probabilities for $q \to
\Lambda(\bkt, \pm \bSt) + X$.}
\label{D1Tperp}
\end{figure}

\begin{figure}[htb] 
\centerline{\epsfig{file=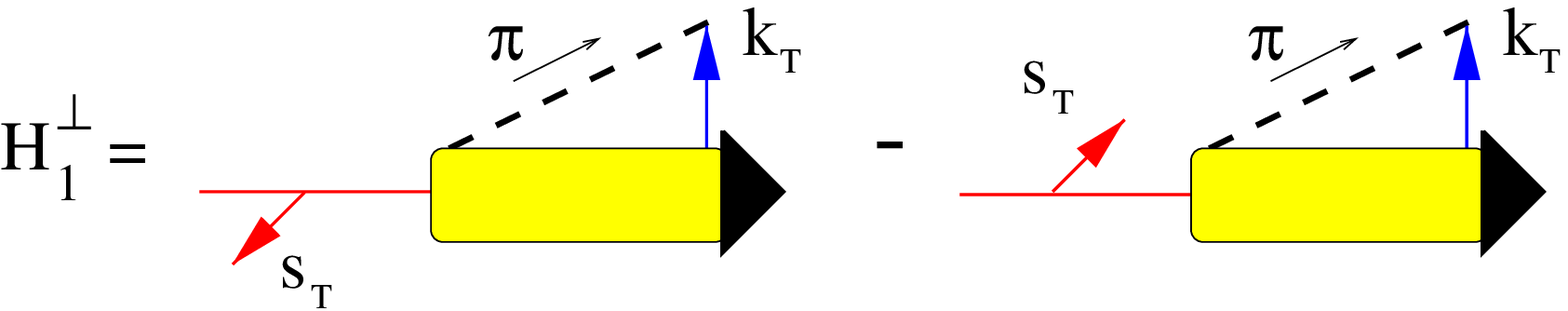,width=4in}}
\vspace{10pt}
\caption{The function
$H_1^\perp$ signals different probabilities for $q(\pm \bSt) \to
\pi(\bkt) + X$.}
\label{H1perp}
\end{figure}

In case there are two large scales ($\sqrt{s}$ and $p_T$) present in 
processes like $p \, p \rightarrow \Lambda^{\uparrow} \, X$ and 
$p \, p^{\uparrow} \rightarrow \pi \, X$ \cite{Adams}, the description 
factorizes into a hard subprocess 
cross section convoluted with two types of soft physics correlation functions.
The latter characterize  
\begin{figure}[htb] 
\centerline{\epsfig{file=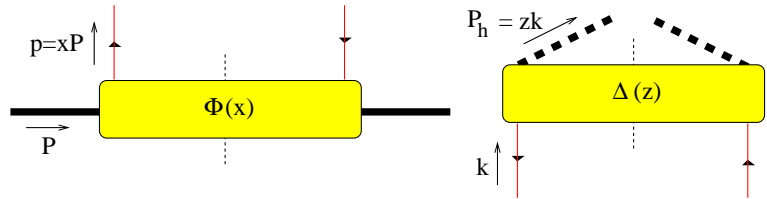,width=4in}}
\vspace{10pt}
\caption{The correlation functions $\Phi(x)$ and $\Delta(z)$.}
\label{phidel}
\end{figure}
the distribution of quarks inside a 
hadron ($\Phi$) and quarks fragmenting into a hadron plus anything ($\Delta$),
see Fig.\ \ref{phidel}.
As a function of the lightcone momentum fraction $x$, $\Phi$ is given by
(after imposing parity and time reversal)
\beq
\Phi(x)=\frac{1}{2} \left[{f_1(x)} \mbox{$\not\! P\,$}+ 
{g_1(x)}\,\lambda\gamma_{5}\mbox{$\not\! P\,$}
+ {h_1(x)}\,\gamma_{5}\mbox{$\not\! S$}_{T}\mbox{$\not\! P\,$}\right].
\label{paramPhi}
\eeq
The transversity function $h_1$ \cite{Ralst-S-79} is the distribution
of transversely polarized quarks inside a transversely polarized hadron. 
Similarly, the fragmentation correlation function 
$\Delta(z)$ \cite{Coll-S-82} is parameterized as
\beq
\Delta(z)=\frac{1}{2} \left[{D_1(z)} \mbox{$\not\! P\,$}+ 
{G_1(z)}\,\lambda\gamma_{5}\mbox{$\not\! P\,$}
+ {H_1(z)}\,\gamma_{5}\mbox{$\not\! S$}_{T}\mbox{$\not\! P\,$}
\right]. 
\eeq
The transversity fragmentation function $H_1$ is the probability that a
transversely polarized quark fragments into a transversely polarized
(spin-1/2) hadron plus anything. The transversity functions $h_1$ and $H_1$ 
are not sufficient to describe single spin asymmetries. 
But if one includes transverse momentum dependence \cite{Ralst-S-79},
then T-odd functions can lead to unsuppressed single spin 
asymmetries (for a detailed explanation cf.\ Ref.\ \cite{Boer-99}). 
The transverse momentum dependent fragmentation functions\footnote{Due to the
problematic nature of T-odd {\em distribution\/} functions \cite{Collins-93b}, 
here we will focus only on T-odd {\em fragmentation\/} functions, 
which are expected to arise due to final 
state interactions, rather than due to time reversal symmetry violation.} 
are defined through
the parameterization of the correlation function $\Delta(z,\bkt)$
\beq
\Delta(z,\bkt)=
\text{T-even part} + \frac{1}{2} \biggl[
D_{1T}^\perp\, \frac{\epsilon_{\mu \nu \rho \sigma}
\gamma^\mu P^\nu k_T^\rho S_{T}^\sigma}{M}
+ H_{1}^\perp\,\frac{\sigma_{\mu \nu} k_T^\mu P^\nu}{M}
\biggr].
\label{Deltaexp}
\eeq
The fragmentation functions $D_{1T}^\perp$ and $H_1^\perp$ are T-odd
functions, linking transverse spin --of a hadron and quark, respectively--
and transverse momentum 
with a specific orientation (handedness), cf.\ Figs.\ \ref{D1Tperp} and
\ref{H1perp}.  
There are a few experimental indications that the ``Collins effect'' function
$H_1^\perp$ \cite{Collins-93b} 
is indeed nonzero \cite{Bravar,HERMES,EST-98}; it can account for a number of 
different pion production asymmetries, including 
$p \, p^\uparrow \to \pi \, X$ \cite{Anselmino}.
The chiral-even function $D_{1T}^\perp$ \cite{Mulders-Tangerman-96} is 
expected to be 
relevant for {transversely polarized $\Lambda$ production} \cite{ABM}, 
e.g.\ in $p \, p \to \Lambda^\uparrow X$. 

\section*{Transversely polarized $\Lambda$ production}

The single transverse spin asymmetry 
\beq
P_{N}= \frac{\sigma(p \, p \to {\Lambda^{\uparrow}} \, X) \, {-} \, 
\sigma(p \, p \to {\Lambda^{\downarrow}} \, X)}{\sigma(p \, p \to
{\Lambda^{\uparrow}}\, X) \, {+} 
\sigma(p \, p \to {\Lambda^{\downarrow}} \, X)} 
\eeq
exhibits similar behavior as
the pion production single spin asymmetries, namely the magnitude grows as a
function of  $p_T^{}$ and $x_F^{}$. The BNL-RHIC collider 
can reveal whether $p \, p \to \Lambda^{\uparrow} \, X$ persists at larger
values of $\sqrt{s}$ and $p_T$, which would be an important indication that 
a factorized picture should be applicable. Such a factorized description 
(requiring $p_T \simorder 1 \; \text{GeV}$) in terms of the quark 
fragmentation function $D_{1T}^\perp$ would imply that the 
production of a transversely polarized $\Lambda$ is independent
of the details of the initial state and unlike existing models of the above 
asymmetry \cite{Felix,Soffer},
one could restrict to the modeling of $D_{1T}^\perp$.
A detailed study of $p \, p \to \Lambda^{\uparrow} \, X$ within the factorized
picture will be presented in Ref.\ \cite{ABM}. 
Here we want to indicate how a fit of $D_{1T}^\perp$ from that data can be 
used to compare to possible future 
$\ell \, p \to \ell' \, \Lambda^{\uparrow} \, X$ data and to 
$e^+ \,e^- \to \Lambda^{\uparrow} \, \text{jet} \, X$ data.

1) Recently, an asymmetry in $e \, p \to \Lambda^{\uparrow} \, X$ was reported 
\cite{Belos} (preliminary): $P_N = 0.066 \pm 0.011 \pm 0.025$ (arising 
mainly from $e \, p \to e' \, \Lambda^{\uparrow} \, X$ events with 
$Q^2 \simeq 0$).
If such an asymmetry would also be found in the semi-inclusive DIS process 
$e \, p \to e' \, \Lambda^{\uparrow} \, X$ (for $Q^2 \simorder 1
\, \text{GeV}^2$) the factorized description \cite{Mulders-Tangerman-96} 
would imply\footnote{The chiral-even function $D_{1T}^\perp$ can also be probed
in {charged current exchange processes}; for the cross section
expressions in semi-inclusive DIS we refer to \cite{BJM-00}.} 
\beq
\frac{d\Delta\sigma(e\, p \to e' \, {\Lambda^\uparrow} \, X)}{
d\sigma(e \, p\to e' \, \Lambda \, X)} = 
\sin(\phi_{S_T^\Lambda}^{}- \phi_{P_T^\Lambda}^{})\; {P_N},
\eeq
where the analyzing power $P_N$ is a function of 
$f_1,D_1$ and ${D_{1T}^{\perp}}$. If one makes an Ansatz inspired by a model
for the Collins function \cite{Collins-93b} ($\eta, M$ are free parameters and
Gaussian transverse momentum dependence of the unpolarized fragmentation
function $D_1$ is assumed):
\beq
{D_{1T}^{\perp}(z,\bm{k}_T^{2})}= {\eta} 
\frac{{M} M_\Lambda}{\bm{k}_T^2 + {M^2}} 
{D_1(z,\bm{k}_T^{2})} \, \, 
\Longrightarrow \, \, 
{P_N} \approx \frac{2 \, {\eta} \, {M} \, Q_T}{Q_T^2 +
4 {M^2}}.
\eeq
This exhibits plausible behavior as a function of the 
transverse momentum ($Q_T$) of the $\Lambda$ and 
this expression for $P_N$ can be used to fit $D_{1T}^\perp$, allowing for a 
check of the universality of ${D_{1T}^{\perp}}$ if compared to the resulting
fit from $p \, p \to \Lambda^{\uparrow} \, X$ data. 

2) In the case of $e^+ \,e^- \to \Lambda^{\uparrow} \, \text{jet} \, X$, one 
determines the transverse momentum ($Q_T$) of the $\Lambda$ compared to the 
jet (or thrust) axis. In the factorized picture with transverse momentum 
dependent functions this will yield a contribution \cite{Boer2}
\beq
\frac{d\sigma({e^+e^-\to \Lambda^{\uparrow} \,\, \text{jet} \,\, 
X})}{d\Omega dz d Q_T^{}}
\propto \sin(\phi_{P_T^\Lambda}^{}- \phi_{S_T^\Lambda}^{})
\frac{Q_T^{}}{M_\Lambda} \sum_{a,\bar a}{e_a^2}\; 
{D_{1T}^{\perp a}(z,z^2 Q_T^2)}. 
\eeq
Note that the exponential fall-off of the function at larger values of $Q_T$
wins out over the explicit power of $Q_T$ (all under the requirement of 
$Q_T^2 \ll Q^2$).
Also we note that for 
${e^+ \,e^- \to} \; {Z} \; {\to \Lambda^{\uparrow} \, \text{jet} \, X}$ 
transverse 
$\Lambda$ polarization has been measured to be small (at the percent level)
\cite{ALEPH-OPAL}. 
We expect 
two effects to contribute to the suppression of the above contribution
at large scales such as $Q=M_Z$. 
The function $D_{1T}^{\perp}$ might be a decreasing function of $Q^2$,
although this is not yet known. Also, 
transverse momentum dependent azimuthal asymmetries suffer from effective 
{\em power\/} suppression due to Sudakov factors \cite{Boer-00}. 
The LEP result would then not contradict a
possibly large asymmetry in $e \, p \to e' \, \Lambda^{\uparrow} \, X$ at
lower energies (measurable at HERMES).   

\section*{Acknowledgements}
I thank M.~Anselmino, U. D'Alesio and F. Murgia for collaboration on this 
topic. Furthermore, I thank RIKEN, Brookhaven National Laboratory and the 
U.S.\ Department of Energy (contract number DE-AC02-98CH10886) for
providing the facilities essential for the completion of this work.


\begin{references}

\bibitem{Lambdadata}
E.g.\ R608 Collaboration, A.M. Smith {\em et al.}, {\it Phys.\ Lett.}\ {\bf
B185}, 209 (1987).

\bibitem{Adams} 
FNAL E704 Collab., {\it Phys.\ Lett.}\ {\bf B261}, 201 (1991); 
{\it Phys.\ Rev.\ Lett.}\ {\bf 77}, 2626 (1996).

\bibitem{Ralst-S-79}
J.P. Ralston and D.E. Soper, {\it Nucl.\ Phys.}\ {\bf B152}, 109 (1979).

\bibitem{Coll-S-82}
J.C. Collins and D.E. Soper, {\it Nucl.\ Phys.}\ {\bf B194}, 445 (1982).

\bibitem{Boer-99}
D. Boer, hep-ph/9912311.

\bibitem{Collins-93b}
J.C. Collins, {\it Nucl.\ Phys.}\ {\bf B396}, 161 (1993).

\bibitem{Bravar}
A. Bravar (for the SMC Collaboration), {\it Nucl.\ Phys.\ (Proc.\ Suppl.)}\ 
{\bf B79}, 520 (1999).

\bibitem{HERMES}
HERMES Collaboration, A. Airapetian {\em et al.}, {\it Phys.\ Rev.\ Lett.}\
{\bf 84}, 4047 (2000).

\bibitem{EST-98}
A.V. Efremov, O.G. Smirnova and L.G. Tkatchev, hep-ph/9812522. 

\bibitem{Anselmino} M. Anselmino, M. Boglione, F. Murgia, {\it Phys.\ Rev.}\ 
{\bf D60}, 054027 (1999);\\
M. Boglione, E. Leader, {\it Phys.\ Rev.}\ {\bf D61}, 114001 (2000).

\bibitem{Mulders-Tangerman-96}
P.J. Mulders and R.D. Tangerman, {\it Nucl.\ Phys.}\ {\bf B461}, 197 (1996).

\bibitem{ABM}
M. Anselmino, D. Boer, U. D'Alesio and F. Murgia, in preparation.

\bibitem{Felix}
J. F\'{e}lix, {\it Mod.\ Phys.\ Lett.} {\bf A14}, 827 (1999). 

\bibitem{Soffer}
J. Soffer, hep-ph/9911373.

\bibitem{Belos}
S.L. Belostotski (for HERMES), 
{\it Nucl.\ Phys.\ (Proc.\ Suppl.)}\ {\bf B79}, 526 (1999).

\bibitem{BJM-00} 
D. Boer, R. Jakob and P.J. Mulders, {\it Nucl.\ Phys.} {\bf B564}, 471 (2000).

\bibitem{Boer2} 
D. Boer, R. Jakob and P.J. Mulders, {\it Nucl.\ Phys.}\ {\bf B504}, 345 (1997);
{\it Phys.\ Lett.}\ {\bf B424}, 143 (1998).

\bibitem{ALEPH-OPAL}
ALEPH Collaboration, {\it Phys.\ Lett.}\ {\bf B374}, 319 (1996);\\
OPAL Collaboration, {\it Eur.\ Phys.\ J.} {\bf C2}, 49 (1998).

\bibitem{Boer-00}
D. Boer, hep-ph/0004217.

\end{references}
\end{document}